\documentstyle[12pt]{article}

\newcommand{\be}{\begin{enumerate}}
\newcommand{\ee}{\end{enumerate}}
\newcommand{\bi}{\begin{itemize}}
\newcommand{\ei}{\end{itemize}}
\newcommand{\bge}{\begin{equation}}
\newcommand{\eeq}{\end{equation}}
\newcommand{\bga}{\begin{eqnarray}}
\newcommand{\eea}{\end{eqnarray}}

\newcommand{\bfx}{{\bf x}}
\newcommand{\bfc}{{\bf c}}
\newcommand{\bfk}{{\bf k}}
\newcommand{\one}{{\bf 1}}
\newcommand{\bfL}{{\bf \Lambda}}
\newcommand{\bfO}{{\bf \Omega}}

\newcommand{\arq}{\langle r\rangle_q}
\newcommand{\rtq}{\langle r^2\rangle_q}

\begin{document}

\title{Lattice gas with ``interaction potential''}
\author{Olivier Tribel and Jean Pierre Boon\\Center for Nonlinear Phenomena
and Complex Systems\\
Universit\'e Libre de Bruxelles - Campus Plaine - CP 231\\
B-1050 Brussels, Belgium.\\
{\small e-mail: {\tt otribel@ulb.ac.be; jpboon@ulb.ac.be}}
}

\date{\today}

\maketitle

\noindent
\begin{abstract}
  We present an extension of a simple automaton model to incorporate
non-local interactions extending over a spatial range in lattice gases.
{}From the viewpoint of Statistical Mechanics, the lattice gas with
interaction range may serve as a prototype for non-ideal gas behavior.
{}From the density fluctuations correlation function, we obtain a quantity
which is identified as a potential of mean force.
Equilibrium and transport properties are computed theoretically and by
numerical simulations to establish the validity of the model at
macroscopic scale.
\end{abstract}
\par
\begin{center}
\end{center}
\section{Lattice gas with non-local interactions}
Standard lattice gas automata (LGA) evolve according to an iterated sequence
of mass- and momentum-preserved local collisions followed by propagation.
Non-local interactions can be incorporated in the LGA dynamics via
long-distance momentum transfer simulating attraction and/or repulsion between
particles \cite{appert}, \cite{GEF}, \cite{tb}. In local collisions, momentum
redistribution is a node-located process with local conservation of mass
and momentum. In non-local interactions (NLI), momentum is exchanged between
two
particles residing on nodes separated by a (fixed or variable) distance $r$:
mass is conserved locally, momentum is conserved globally. At the
macroscopic level, the main feature exhibited by LGA models with NLI's is
a  ``liquid-gas''-type phase separation with bubble and drop formation
\cite{appert}. From the statistical mechanical viewpoint, LGA's with NLI's
form an interesting class of models in that --- in contrast to standard
collision-propagation models --- they include an elementary process which
is essential for ``non-ideal'' behavior.

The dynamics of LGA virtual particles is not governed by Newton's equation
of motion and the concepts of force and potential cannot be used in the sense
of
classical mechanics. Moreover in real fluids each particle is subjected
{\em a priori} to the force field of {\em all} particles (whose effect is
quantified by the potential of mean force) whereas in discrete lattice gases
each particle interacts non-locally with at most {\em one} other particle at
a time. So {\em stricto sensu} the usual concept of intermolecular
potential does not apply to lattice gases.

In the LGA model with NLI's proposed by Tribel and Boon \cite{tb}, the idea
of an interaction range was introduced by governing the interaction distance
according to a probability distribution --- namely a power law
$(\propto r^{-\mu})$ --- wherefrom a distance $r$ is drawn for each particle
at every time step. Here we show that for sufficiently long times and large
number of particles, the implementation of a probability distribution of
interaction distances has a resulting effect similar to the effect of an
interaction potential.
We first describe the model in section \ref{model}. Then in section
\ref{correlations} we compute the density fluctuation correlations \cite{corr}
wherefrom
a quantity is obtained which can be identified as a potential of mean force.
 Sections \ref{equilibrium} and
\ref{transport} present an analysis of the equilibrium and transport
properties.
We conclude with some comments.

\section{Interaction range model}\label{model}
The automaton resides on a two-dimensional triangular lattice and uses for
propagation and local collisions the rules of the FHP-III model \cite{dl}
with periodic boundary conditions. Non-local interactions can take place
 between two particles when nodes separated by some distance $r$ exhibit
favorable configurations as illustrated in Fig.1.
The interaction
modifies the orientation of the velocity vectors from a diverging configuration
to a converging configuration to simulate attractive forces and {\it
vice-versa}
for repulsive forces. At each time step, the algorithmic procedure must realize
a pairing of particles separated by a distance $r$ drawn from a probability
distribution $p(r)$. It is clear that a parallel algorithm can hardly be
efficient here. Therefore we use a sequential algorithm which proceeds as
follows:
\be
\item at each time step, a direction is arbitrarily chosen along any of the
lattice axes and all interactions will be along that direction during that
time step;
\item a particle, say at node $A$, is (sequentially) selected and accepted
if its state has not been modified by a previous interaction in the sequential
procedure;
\item a distance $r$ is drawn from the distribution $p(r)$ and a pointer is
set at nodes $F$ and $B$, located respectively at a distance $r$ forward and
backward from $A$;
\item if one of the configurations ``$BA$'' or ``$AF$'' is compatible for
interaction (see Fig.1), the configuration is modified
accordingly and the procedure keeps
track of the modification for the duration of the sequence (each particle
can undergo no more than one interaction per time step).
\ee
As a result the effective probability that an interaction occurs in the
simulation differs from the theoretical $p(r)$. The details of the computation
are given in the Appendix; here we merely quote the final result which
expresses the {\it effective probability}
 $q(r)$ in terms of $p_F(r)$ and $p_B(r)$,
denoting respectively the forward and backward probabilities with the imposed
analytical form
(e.g. $p(r)\propto r^{-\mu}$):
\bge\label{eq:qr}
q(r) = p_F(r)+p_B(r)-p_F(r)p_B(r)
\eeq
with
\bge\label{eq:pFr}
p_F(r) = p(r)\prod_{\ell=r+1}^{r_{max}}\left[1-\kappa_2p_F(\ell)\right]
\eeq
and
\bge\label{eq:pBr}
p_B(r)=p(r)\left\{\prod_{\ell=1}^{r_{max}}\left[1-\kappa_2p_F(\ell)\right]
\right\}^2\prod_{\ell=1}^{r-1}\left[1-\kappa_2p_B(\ell)\right]
\eeq
where $r_{max}$ is the cutoff distance
in the distribution $p(r)$ (see \cite{tb}) and
$\kappa_2=f(1-f)$, with
$f$ the particle density per channel\footnote{Note that if desired
operationally, Eqs. (\ref{eq:qr})--(\ref{eq:pBr}) can be inverted numerically
to obtain a function $p(r)$ such that the effective distribution $q(r)$ be of
given analytical form.}. Besides the fact that here interaction distances are
distributed over an interaction range, an important difference with the
fixed-distance model (\cite{appert}, \cite{GEF}) is that in the present case,
each particle belongs to many pairs of possible interactions. Note also that
for similar reasons, there is a bias in the effective distribution toward large
interaction distances. Indeed when drawing a low value of $r$ from $p(r)$ in
the
sequential procedure, there is a greater chance that the second particle of the
pair be already involved in a previous pairing. As a result the effective
probability for long distance interactions is larger than predicted by the
pre-set distribution $p(r)$.

The question now arises as to define a quantity which can be identified as an
interaction potential in a discrete lattice gas. We propose the following
heuristic argument. We evaluate the rate of momentum exchange caused by
the non-local interaction
\bge\label{eq:Fr}
F(r) = \gamma\kappa_2^2q(r)
\eeq
where $\gamma $ is a numerical factor whose value corresponds to the average
amount of momentum transfer
($\gamma = 4/3$ and $\gamma =1$ for the models shown in Figs.
1.a and 1.b respectively).
Interpreting $F(r)$ in (\ref{eq:Fr}) as a force, we define the ``pair
potential'' as the discrete analogue of the potential in continuum mechanics:
\bga\label{eq:ur}
u(r) & = &  -\kappa_2^{-2} \sum_{\ell=1}^r F(\ell) \nonumber\\
& = & -\sum_\ell \gamma q(\ell) \nonumber\\
& = & - \gamma {\cal F}(r)
\eea
where ${\cal F}(r)$ is the repartition function corresponding to the
distribution $q(r)$. Then using Eqs. (\ref{eq:qr})--(\ref{eq:pBr}),
$u(r)$ is well-defined once $p(r)$ is fixed. For instance, if we use the
power-law distribution $p(r)\propto r^{-\mu}$ such that the interactions are
repulsive for $r=1$ and attractive for $r=2,\ldots,r_{max}$,
$u(r)$ exhibits a form compatible
with the expected typical pair interaction potential, as shown in Fig.2.

\section{Density fluctuation correlations}\label{correlations}
The next question is the influence of the non-local interactions on the density
fluctuation correlations in the lattice gas, which is most conveniently
measured by the static structure factor \cite{pg} defined by
\bge\label{eq:Sk}
\rho S(k) = \frac{1}{T}\sum_{t=1}^T\sum_{i,j}\delta n_i^*(k,t)\delta
n_j(k,t),
\eeq
where $\rho$ is the density per node, and
\bge\label{eq:fi}
\delta n_i(k,t) = \sum_{\bf x}e^{-\imath{\bf k}\cdot{\bf x}}\left[n
({\bf x},{\bf c}_i;t)-f\right]
\eeq
is the fluctuation of the channel occupation number $n_i$ ($i=0\ldots b$).
In the ideal lattice gas (whose dynamics is governed by propagation-collision
rules) there are no static density correlations and the static structure
factor is a constant \cite{pg}:
\bge\label{eq:S0k}
S^0(k) = (1-f)(1-\delta(k)).
\eeq
By analogy with the statistical mechanical theory of continuous fluids
\cite{by}, we write
\bge\label{eq:bysk}
\frac{S(k)}{S^0(k)}=1+fh(k)
\eeq
where $h(k)$ is the Fourier transform of the pair correlation function
$[g(r)-1]$
and is therefore related to the potential of mean force $\phi(r)$ since
$g(r)=exp[-\beta\phi(r)]$ (here $\beta$ is an arbitrary constant).
So by measuring the density fluctuation correlations
in lattice gas simulations, we can extract a function $\phi(r)$ from the
measured static structure factor. The results are shown in Fig.3:
both the radial distribution
function $g(r)$ and the potential function $\phi(r)$ are
reminiscent of those obtained in real fluid measurements \cite{egelstaff}.
The connection between the potential of mean force $\phi(r)$ and the
interaction
potential $u(r)$ discussed in section \ref{model} remains to be clarified.

Another interesting feature is worth mentioning. Consider the LGA is in the
appropriate density range for spinodal decomposition (see section
\ref{equilibrium}). Then one could effectively ``quench'' the system by
increasing the interaction range.
By measuring $S(k)$ at successively increasing values of $r_{max}$ we find
that $S(k)$ increases dramatically at low $k$. Following the lines of heuristic
reasoning and anticipating a result of section \ref{equilibrium}, we infer that
\bge\label{eq:Sk->0}
S(k\rightarrow 0)\simeq \frac{1-f}{1- \gamma \kappa_3\langle r\rangle_q}
\eeq
where the denominator follows from the expression for the compressibility
(see section \ref{equilibrium}). Here $\langle r\rangle_q$ is the expectation
of $r$ computed with the distribution $q(r)$. Since $\langle r\rangle_q$
increases with $r_{max}$, $S(k\rightarrow 0)$ grows accordingly as expected
when the phase transition is approached. Further analysis will be presented
in a forthcoming paper.

\section{Equilibrium properties}\label{equilibrium}

The pressure at global equilibrium is given by (see {\cite{feynman})
\bge\label{eq:defp}
p = \frac{1}{2V}\sum_{\bfx\in{\cal L}}\sum_i
\langle n_i(\bfx) + m_i(\bfx) \rangle,
\eeq
where $V$ is the number of nodes of the automaton universe,
$\sum_{\bfx\in{\cal L}}\sum_i
\langle n_i(\bfx)\rangle$ is the momentum transport due
to propagation and $\sum_{\bfx\in{\cal L}}\sum_i
\langle m_i(\bfx)\rangle$ is the momentum
flux due to NLI's. Then the hydrostatic pressure can be evaluated as follows:
\bi
\item the convective momentum flux is the total momentum carried by moving
particles in the fluid. On each node in the FHP-III model, there are 6 channels
with velocity $1$ and one zero-velocity channel, so that
\bge
\sum_{\bfx\in{\cal L}}\sum_i
\langle n_i(\bfx)\rangle = V\frac{6}{7}\rho,
\eeq
where $\rho$ is the average density per node;
\item the non-local momentum flux is caused by NLI's. The value of this
flux is clearly given by
\bge
\sum_{\bfx\in{\cal L}}\sum_i
\langle m_i(\bfx)\rangle = 3V \gamma \kappa_2^2\sum_rrq(r),
\eeq
since on the average each NLI causes a momentum flux of value $\gamma r$ and
on a given node $3$ particles may independently be involved in an interaction.

\ei
Consequently
\bge\label{eq:pressure}
p = 3f-\frac{3}{2} \gamma \,\kappa_2^2\arq
\eeq
with $f = \frac{\rho}{7}$ and $\arq = \sum_rrq(r)$. The first term on
the r.h.s. of (\ref{eq:pressure}) is the kinetic pressure of the ideal
lattice gas and the second term as given in this
mean-field evaluation depends only on the first moment of the
distance distribution $q(r)$.
Note that for fixed-distance interaction models, $q(r) = \delta(r-\ell)$ and
(\ref{eq:pressure}) becomes for the model of Fig.1.a
(with $\gamma = \frac{4}{3}$)
\bge
p = 3f - 2\ell\kappa_2^2
\eeq
as given in \cite{appert} and \cite{GEF}.

\par
{}From (\ref{eq:pressure}) it follows that the compressibility is given by
\bga\label{eq:defchi}
\chi & = & \frac{1}{\rho}\frac{\partial\rho}{\partial p}\nonumber\\
& = & \frac{1}{\frac{3}{7}\rho\left(1-\gamma \arq\kappa_3\right)}\nonumber\\
& = & \frac{\chi^0}{1- \gamma \arq\kappa_3},
\eea
where $\kappa_3 \equiv f(1-f)(1-2f)$ and $\chi^0$ is the compressibility of
the ideal gas.
By using $S(k\rightarrow~0~) = \rho\beta^{-1}\chi _{\it th}$ \cite{by} and the
{\it thermodynamic} pressure of the ideal gas
$p_{\it th} = -b \beta^{-1}$ln$(1-f)$ which yields the compressibility
$\chi^0 _{\it th} = \rho^{-1} \beta (1-f)$ (with $\beta^{-1} = c_0^2 = 3/7$ for
the FHP-III model) we obtain Eq.(\ref{eq:Sk->0}).

The compressibility equation (\ref{eq:pressure}) yields the square of the
sound velocity $c_s$
\bge\label{eq:cs}
c_s^2 = \frac{\partial p}{\partial \rho}
=c_0^2\left[1-\gamma \kappa_3\arq\right].
\eeq

Eq. (\ref{eq:cs}) is valid as long as
$\frac{\partial p}{\partial \rho} > 0$; when $\frac{\partial p}{\partial\rho} <
0$, the density fluctuations show an explosive behavior
and the system separates into two phases (i.e. for $\arq > 7.79$).
In Fig.4 we show the results of measurements
of the pressure as given by Eq. (\ref{eq:defp}), compared to the theoretical
prediction (\ref{eq:pressure}); in Fig.5, the theoretical
sound velocity (\ref{eq:cs}) is compared to the simulation results (for the
experimental method, see e.g. \cite{GEF}). Experimental evidence of spinodal
decomposition is given in Fig.6 where the evolution of the
density distribution shows how phase separation takes place in the automaton.

\section{Transport coefficients}\label{transport}

\subsection{Microdynamical and lattice Boltzmann equations}
The evolution of the automaton is obtained by applying successively the
non-local interaction routine, the collision routine, and the propagation
routine. This computational procedure is the operational realization of the
microscopic dynamics of the automaton whose mathematical formulation is
given by the microdynamical equation
\bge\label{eq:microdynamical}
n(\bfx+\bfc_i,\bfc_i;t+1) = {\cal C}_i\left\{{\cal I}\left[
n(\bfx;t)\right]\right\}
\eeq
where $n(\bfx,\bfc_i;t)$ is the Boolean occupation variable of channel $i$
at node $\bfx$ at time $t$. ${\cal C}$ and ${\cal I}$ are  the local
collision and non-local interaction operators respectively. The explicit
expression of the non-local operator ${\cal I}$ reads
\bge
{\cal I}_i = \frac{1}{3}\sum_rq(r)\sum_{j=-1}^{+1}{\cal I}^{\,r}_{i,i+j},
\eeq
with (for the model of Fig.1.a)\footnote{Channel indices are numbered
counter-clockwise from 0 to 5 for moving particles and 6 for the rest particle,
 and indices $i$ and $i+j$ are taken modulo 6.}

\bga\label{eq:defI}
{\cal I}^{\,r}_{i,i}n(\bfx;t) & = &
\left[\overline{n}_i(\bfx;t)n_{i+3}(\bfx;t)\right]
\left[n_i(\bfx+r\bfc_i;t)\overline{n}_{i+3}(\bfx+r\bfc_i;t)\right] \nonumber\\
& - & \left[n_i(\bfx;t)\overline{n}_{i+3}(\bfx;t)\right]
\left[\overline{n}_i(\bfx-r\bfc_i;t)n_{i+3}(\bfx-r\bfc_i;t)\right] \nonumber\\
{\cal I}^{\,r}_{i,i\pm 1}n(\bfx;t) & = &
\left[\overline{n}_i(\bfx;t)n_{i\mp 1}(\bfx;t)\right]
\left[n_i(\bfx+r\bfc_{i\pm 1};t)\overline{n}_{i\mp 1}(\bfx+r\bfc_{i\pm 1};t)
\right] \nonumber\\
& - & \left[n_i(\bfx;t)\overline{n}_{i\mp 1}(\bfx;t)\right]
\left[\overline{n}_i(\bfx-r\bfc_{i\pm 1})n_{i\mp 1}(\bfx-r\bfc_{i\pm 1};t)
\right],
\eea
where $\overline{n} \equiv 1-n$.

Taking the average of Eq. (\ref{eq:microdynamical}) over a
non-equilibrium ensemble, and making the {\it molecular chaos} assumption,
one obtains the lattice Boltzmann equation
\bge\label{eq:Boltzmann}
f(\bfx+\bfc_i,\bfc_i;t+1) = {\cal C}_i\left\{{\cal I}\left[f(\bfx;t)\right]
\right\},
\eeq
where $f(\bfx,\bfc_i;t) \equiv \langle n(\bfx,\bfc_i;t)\rangle$ is the singlet
distribution function of channel $i$ at node $\bfx$ at time $t$.
In this equation, the operators ${\cal C}$ and ${\cal I}$ act on the
distribution function $f$ (not on the Boolean variables $n_i$).

\subsection{Linearized lattice Boltzmann equation}
Considering small deviations from local equilibrium
\bge
f(\bfx,\bfc_i;t)=f+\delta f(\bfx,\bfc_i;t)
\eeq
the lattice Boltzmann equation (\ref{eq:Boltzmann}) may be linearized for
the  perturbation $\delta f$. Denoting by $\bfO$ the usual linearized
collision operator and by $\bfL$ the linearized NLI operator
\bge\label{def:Lambda}
\delta f'(\bfx,\bfc_i;t) = (\one + \bfL)_{ij}\delta f(\bfx,\bfc;t),
\eeq
the linearized lattice Boltzmann equation reads
\bge\label{eq:LLBE}
\delta f(\bfx+\bfc_i;\bfc_i;t+1) = (\one+\bfO)_{ij}(\one+\bfL)_{jk}\delta f
(\bfx,\bfc_k;t).
\eeq

We develop the perturbations $\delta f$ in Fourier modes
\bge
\delta f(\bfx, \bfc_i;t) =
\sum_\mu\sum_\bfk\psi_\mu(\bfk,\bfc_i)e^{\imath\bfk\cdot\bfx+z_\mu(\bfk)t},
\eeq
and rewrite Eq.(\ref{eq:LLBE}) in Fourier space to obtain the eigenvalue
equation for the automaton
\bge\label{eq:eigen}
\left[e^{z_\mu(\bfk)+\imath\bfk\cdot\bfc}-
(\one+\bfO)(\one+\bfL)\right]_{ij}|\psi_\mu(\bfk,\bfc_j)\rangle = 0.
\eeq
This equation is formally identical to the eigenvalue equation for the
fixed-distance model \cite{GEF}, but the operator $\bfL$ is now a linear
combination
of the corresponding fixed-distance operators. The three slow (hydrodynamic)
modes of interest are the shear mode, noted $\psi_\nu$, whose
eigenvalue corresponds to the kinematic viscosity $\nu$, and
the two sound modes $\psi_{\sigma = \pm}$ related to the sound velocity $c_s$
and damping coefficient $\Gamma$.

\subsection{Transport coefficients}
As mentioned, the operator $\bfL$ is a linear combination of fixed-distance NLI
operators
\bge
\bfL_{ij} = \sum_rq(r)\Lambda_{ij}^{*r}
\eeq
where $\Lambda_{ij}^{*r}$ denotes the NLI operators at distance $r$.
So the results for the fixed-distance interaction operators can be extended to
the present model (see \cite{GEF} for details). The main difference is that
the distance
$r$ is now replaced by its mean value $\arq$, and $r^2$ by the variance
 $\rtq = \sum_r\;r^2q(r)$. Expanding the eigenvalues in powers of $k$
\bge
z_\mu(\bfk) = (\imath k)z_\mu^{(1)} + (\imath k)^2z_\mu^{(2)} + \ldots,
\eeq
we obtain
\bge
\begin{array}{cc}
z_\sigma^{(1)} = \pm c_s, & z_\nu^{(1)} = 0, \nonumber\\
z_\sigma^{(2)} = \Gamma = \frac{1}{2}(\nu+\zeta), &
z_\nu^{(2)} = \nu,
\end{array}
\eeq
with
\bga\label{eq:transport}
\nu & = &
\nu_0\left(1-\frac{1}{3}\arq\kappa_3\right)\left(1-\arq\kappa_3\right)
+ \frac{1}{12}\arq\kappa_3\left(1-\frac{1}{2}\arq\kappa_3\right) +
\frac{1}{8}\rtq\kappa_2, \nonumber\\
\zeta & = & \zeta_0\left(1-\frac{4}{3}\arq\kappa_3\right)-
\frac{1}{21}\arq\kappa_3+\frac{1}{4}\rtq\kappa_2.
\eea
$\nu_0$ et $\zeta_0$ are the kinematic and bulk viscosities of the
standard FHP-III model
\bga\label{eq:no0zeta0}
\nu_0 & = & \frac{1}{14}\left(\frac{1}{\omega_\nu}-\frac{1}{2}\right),
\nonumber\\
\zeta_0 & = & \frac{1}{14}\left(\frac{1}{\omega_\zeta}-\frac{1}{2}\right),
\eea
with
\bga
\omega_\nu & = & \kappa_2(7-8\kappa_2), \nonumber\\
\omega_\zeta & = & 7\kappa_2(1-2\kappa_2).
\eea
Excellent agreement is obtained between the simulation data and the
theoretical results as shown in Fig.7.

\section{Comments}

The question was raised by Gerits {\it et al.} \cite {GEF} that models with
non-local fixed interaction distance lack detailed balance and that therefore
their equilibrium distribution is not known. In the model presented here the
constaint is weaker in that interaction distances are distributed and the
non-local interactions are probabilistic. Yet the condition for semi-detailed
balance was not taken into account. The mean field theory is found to predict
correctly macroscopic equilibrium and transport properties. Fluctuation
correlations were measured and the static structure factor was used to extract
the LGA analog of a potential of mean force. The statistical mechanical theory
for the static and
dynamic structure factors will be discussed in a forthcoming
paper.

\appendix
\section{Evaluation of $q(r)$}
Here we show how to compute the effective automaton distance distribution
$q(r)$ from a given probability distribution $p(r)$.
The algorithmic procedure considers each node sequentially. Define the
node currently examined as the ``center node'' $A$ and define the ``forward
node'' $F$ and ``backward node'' $B$ located  along the direction of
interaction at a distance $r$ on each side of $A$. Now each particle on $A$
may interact with
at most one particle located {\em either} on $F$ {\em or} on $B$. However
since the algorithm is sequential, the forward and backward probabilities are
different: a backward interaction is
possible only if the particle on $A$ has not interacted before, while a
forward interaction is independent of previous interactions involving
node $A$.

\bi
\item {\it Forward interaction probability} : $p_F(r)$. Suppose that
configurations
on $A$ and $F$ are favorable (see Fig. 1). Then the only
additional condition is that the particle on $F$ cannot have been involved
previously in an interaction from a distance larger than $r$. This
``non-event''
 has the probability
\bge
\prod_{\ell = r+1}^{r_{max}}\left[1-\kappa_2p_F(\ell)\right].
\eeq
Consequently the equation for $p_F$ is given by
\bge\label{eq:pF}
p_F(r) = p(r)\prod_{\ell = r+1}^{r_{max}}\left[1-\kappa_2p_F(\ell)\right].
\eeq
\item {\it Backward interaction probability} : $p_B(r)$. We must consider
that (i) the particle on $A$ has not been involved in a previous (forward)
interaction, (ii) the particle on $B$ has not been paired successfully
with another particle when $B$ was a center node (forward interaction),
and (iii) the particle on $B$ has not been paired successfully with a center
node located between $B$ and $A$ (backward interaction).
The corresponding probabilities are:
\be
\item no interaction with $A$ as a forward node:
\bge\label{eq:pB1}
\prod_{\ell = 1}^{r_{max}}\left[1-\kappa_2p_F(\ell)\right];
\eeq
\item no interaction with $B$ as a center node:
same as (\ref{eq:pB1});
\item no interaction with $B$ as backward node:
\bge
\prod_{\ell=1}^{r-1}\left[1-\kappa_2p_B(\ell)\right].
\eeq
\ee
The backward probability is therefore
\bge\label{eq:pB}
p_B(r) = p(r)\left\{\prod_{\ell = 1}^{r_{max}}
\left[1-\kappa_2p_F(\ell)\right]
\right\}^2\prod_{\ell=1}^{r-1}\left[1-\kappa_2p_B(\ell)\right].
\eeq
\ei
As a result $q(r)$ is expressed as a combination of $p_B$ and $p_F$,
given that forward and backward interactions are mutually exclusive:
\bge
q(r) = p_F(r) + p_B(r)[1 - p_F(r)].
\eeq
In the case of fixed-distance interactions (at a distance $\ell _0$),
the above considerations do not apply and the distribution reduces to
\bge
q(r) = \delta_{r,\ell_0}.
\eeq

\bigskip

\begin{center}
{\bf Acknowledgements}
\end{center}
OT has benefited from a grant from the Institut pour l'Encouragement de la
Recherche Scientifique dans l'Industrie et l'Agriculture (IRSIA,Belgium).
JPB acknowledges support
from the Fonds National de la Recherche Scientifique (FNRS, Belgium).

\newpage
{\Large \bf Figure Captions.}
\medskip

Figure 1: Interaction configurations: illustration of configuration changes
through non-local interactions between pairs of particles on nodes at distance
$r$ from each other. Dotted (full) arrows indicate channel occupation before
(after) interaction: $[Q_j,X_k] \rightarrow [Q_k, X_j]$, $X=R,S,T$ where
the channel indices $(j,k)$ are given modulo 6 for $i = 0,\ldots,5$.
Momentum exchange through interactions is two or four units for (a)
configurations whereas all (b) interactions exchange two momentum units.

\par
Figure 2: Pair interaction potential: $u(r)$ for  a
power-law distribution $p(r)\propto r^{-\mu}$ for $1 \leq r \leq r_{max}$,
with $\mu = 1.2$ and $r_{max}=10$. Potential units are arbitrary.

\par
Figure 3: (a) Radial distribution function $g(r)$.
 (b) Potential function $\Phi(r) \equiv \ln(g(r))$.
Probability distribution
$p(r) \propto r^{-\mu}$ for $1 \leq r \leq r_{max}$;
$r_{max} = 6$, $\mu = 0$ (circles),
$r_{max} = 8$, $\mu = 0$ (squares),
$r_{max} = 10$, $\mu = 1$ (diamonds).
Lines are guides to the eye.
Mean channel density $f = 0.1$; lattice size 512x512;
$g(r)$ measured over 500 time steps.

\par
Figure 4: Pressure versus density. (a): fixed-distance interactions; (b):
distributed-distance interactions, with ``flat'' distribution $p(r) \propto
r_{max}^{-1}$. Symbols are experimental data, curves are
theoretical predictions.
Fixed distances $r$ and cutoff distances $r_{max}$ are as indicated.
Lattice size 512x512; each point obtained by averaging over 300 time steps.

\par
Figure 5: Sound velocity versus density. Circles, squares, diamonds
(resp. full, dotted, and dashed lines) correspond to fixed-distance
interactions; triangles (long-dashed line) correspond
to ``flat'' distributed-distance
interactions. Symbols are experimental data, curves are theoretical
predictions. Fixed distances $r$ and cutoff distances $r_{max}$ are as
indicated.
Lattice size 512x512; each point is obtained by averaging over 10 runs of
2500 time steps.

\par
Figure 6: Evolution of the density distribution, measured every 100 time steps
for a total duration of 700 time steps.
The evolution shows horizontal
separation of density peaks characteristic of spinodal decomposition,
as opposed to vertical growth of peaks
in  nucleation and growth processes. Lattice size 512x512; $\mu= 0$;
$r_{max} = 20$; density values are coarse-grained by averaging over each
node and its six nearest neighbors.

\par
Figure 7: (a) Kinematic viscosity $\nu$ versus density;
(b) Sound damping coefficient $\Gamma$ versus density. Symbols are
experimental data, curves are theoretical predictions, conditions are
as indicated. Lattice size 512x512; each point
obtained by averaging over 100 runs of 2500 time steps.

\end{document}